\documentclass{JHEP3}

\newcommand{\be}{\begin{equation}}
\newcommand{\ee}{\end{equation}}
\newcommand{\bea}{\begin{eqnarray}}
\newcommand{\cc}{{\cal C}}
\newcommand{\ck}{{\cal K}}

\newcommand{\cl}{{\cal L}}
\newcommand{\da}{\dagger}
\newcommand{\ddz}{\frac{d}{dz}}
\newcommand{\ddzt}{\frac{d^2}{dz^2}}

\newcommand{\ea}{\end{array}}
\newcommand{\eea}{\end{eqnarray}}
\newcommand{\fr}{\frac}
\newcommand{\cp}{{\cal P}}

\newcommand{\lb}{\label}
\newcommand{\ld}{\ldots}
\newcommand{\lcb}{\left\{}

\newcommand{\lmd}{\left|}
\newcommand{\lra}{\longrightarrow}
\newcommand{\lrb}{\left(}
\newcommand{\lsb}{\left[}
\newcommand{\nn}{\nonumber}

\newcommand{\rra}{\right\rangle}
\newcommand{\rcb}{\right\}}
\newcommand{\rrb}{\right)}
\newcommand{\ba}{\begin{array}}
\newcommand{\rsb}{\right]}

\newcommand{\half}{\fr{1}{2}}

\title{POLYNOMIAL ALGEBRAS AND THEIR APPLICATIONS}
\author{Bindu A. Bambah, School of Physics, University of Hyderabad,
Hyderabad -500 046, INDIA.\\ Invited talk at "The International
conference on Supersymmetric Quantum Mechanics ,\\ Valladolid
Spain (July 15-10, 2003}
 \abstract{A way to construct and classify
the three dimensional polynomially deformed algebras is given and
the irreducible representations is presented. for the quadratic
algebras 4 different algebras are obtained and for cubic algebras
12 different classes are constructed. Applications to quantum
mechanical systems including supersymmetric quantum mechanics are
discussed }

\begin{document}

\section{ Introduction}
 Non linear algebras have been  useful in describing  the algebraic
structure of many physical systems during the last decade. The
q-deformation of the Lie algebras, the quantum groups and $W$
algebras have been studied extensively both in physics as well as
in mathematics\cite{h,s,Ga,G,Le}. They are infinite dimensional
algebras belonging to the class of universal enveloping algebras
of standard Lie algebras. Besides these, algebras which are finite
polynomial deformations of the Lie algebras  , known as non linear
polynomial algebras  have also been observed as the dynamical
symmetry algebras and  as  spectrum generating algebras of many
quantum mechanical systems . In  a large class of these systems
the spectra are not linear ,
  but quadratic or of higher orders
 and if the Casimir operator is taken as
  a Hamiltonian (as in most cases), the advantages
   of using Lie algebras as a dynamical symmetries are lost.
To use the dynamical symmetry properties effectively,
 it is necessary to extend
  the Lie algebra to polynomial algebras
   in which the Hamiltonian is one of the diagonal generators.
 In this way,
  one can relate the degree of degeneracy
   to the dimensions of the representation
   and find out how the additional symmetry
    generators transform one degenerate
 eigenstate to another.
  For this it is useful also that additional symmetry generators
  close on a finite algebra, which in most cases is a quadratic algebra.

Polynomial Algebras with three generators $ N_i , N_j,N_k$ are
defined by the relations
\be
\lsb N_i , N_j \rsb = C_{ij} \lrb N_k \rrb\,, \lb{cr} \nn \ee
where the functions $C_{ij}$ are polynomial functions of the
generators $\lcb N_k \rcb$constrained by the Jacobi identity
\be
\lsb N_i , C_{jk} \rsb + \lsb N_j , C_{ki} \rsb + \lsb N_k ,
C_{ij} \rsb = 0\,. \nn \ee

 In many cases these non-linear algebras admit a coset structure
  i.e among the generators there is a linear subalgebra and
   the commutator of the remaining generators give a symmetric function of the generators of the linear algebra.
$p$ of the $d$ generators $N_i$ satisfy
\begin{equation}
\left[ N_i ,N_j \right] = f^{k}_{ij} N_k , ~~~~~~~i,j=1...p.
\end{equation}
The remaining $d-p$ generators satisfy,
\begin{eqnarray}
\left[ N_i ,N_{\alpha} \right]& = &f^{\beta}_{i \alpha}N_{\beta} ,
\nonumber \\ \left[ N_{\alpha} ,N_{\beta} \right] &=& f_{\alpha
\beta} (N_{\gamma}) ,
\end{eqnarray}

A simplest example of such algebra is the case with $d=1$.
\begin{eqnarray}
\left[ N_0 ,N_{\pm} \right] &=& \pm N_{\pm} \nonumber \\ \left[
N_+ ,N_{-} \right] &=& f(N_0)
\end{eqnarray}
This is also known as polynomial $SU(2)$ or $SU(1,1)$ algebra, a
 three
dimensional polynomial algebra.
 A Casimir operator for this algebra can be taken as
 \be
C=N_- N_+ + g(N_0 ) = N_+ N_- + g(N_0 -1)\nn \ee \be f(N_0)=
g(N_0) -g(N_0 -1) \nn \ee where $g(N_0 ) $is a polynomial in $N_0
$having one degree higher
 than $f(N_0 )$

Three dimensional quadratic algebras are defined by the relations
\bea \left[ Q_0, Q_{\pm} \right] &=& \pm Q_{\pm} \nonumber \\
\left[ Q_+, Q_- \right] &=& a Q_{0}^{2} + b Q_0 + c ,\nn
 \eea
 The Casimir operator for the algebra  is given by,
\bea \cc &=& Q_+ Q_- + g(Q_0 -1) \nonumber \\ \cc &=& Q_+Q_- +
\fr{1}{3} a Q_0^3 - \half(a-b)Q_0^2
             + \fr{1}{6}(a-3b+6c)Q_0 - c\,\nn\eea
 A
general three dimensional cubic algebra  with a coset structure is
given by, \bea \left[ C_0, C_{\pm} \right] &=& \pm C_{\pm}
\nonumber \\ \left[ C_+, C_- \right] &=& a C_{0}^{3} + b C_{0}^{2}
+ cC_{0} +d ,\nn \eea The structure constants $a,b,c$ and $d$ are
constants. The Casimir operator for this algebra is given by \bea
\cc &=& C_- C_+ + \fr{a}{4} C_{0}^{4} + (\fr{b}{3} - \fr{a}{2}
)C_{0}^{3} + (\fr{a}{4} + \fr{b}{2} + \fr{c}{2} )C_{0}^{2} +
(\fr{b}{6} \nn \\&-&\fr{c}{2} +d)C_0 \nn \\ \eea
An  $n^{th}$
order three dimensional polynomial algebra, generated by
$\cp^{n}_+$, $\cp^{(n)}_- $ and $\cp^{(n)}_0$ is defined as \bea
\lsb \cp^{(n)}_0 , \cp^{(n)}_{\pm} \rsb &=&  \cp^{(n)}_{\pm} \nn
\\ \lsb \cp^{(n)}_+ , \cp^{(n)}_- \rsb &=& f_{n} (\cp^{(n)}_0 )
\lb{4.0.1} \eea Here $f_{n} (\cp_0 )$ is an $n^{th}$ order
polynomial in $\cp^{(n)}_0$.

\section{Jordan Schwinger Like Construction Of Polynomial Algebra}

 Let $(\cp^{(n)}_{\pm}, \cp^{(n)}_0 )$ and $(\cp^{(m)}_{\pm} ,
\cp^{(m)}_0)$ be the two sets of two dimensional polynomial
algebras of order $n$ and $m$ respectively. Then the following
operators defined as

\bea \Pi_+ &=& \cp^{(m)}_+ \cp^{(n)}_+ \,\,\,,\,\,\, \Pi_- =
\cp^{(m)}_- \cp^{(n)}_- ,\nn \\ \Pi_0 &=&  \fr{\cp^{(m)}_0
+\cp^{(n)}_0 }{2} \eea
 will always satisfy a polynomial algebra of order
$m+n+1$ if $\Pi   =  \fr{\cp^{(m)}_0 -\cp^{(n)}_0 }{2}$ is a
constant of motion.\\ Similarly \bea \Pi_+ &=& \cp^{(m)}_+
\cp^{(n)}_- \,\,\,,\,\,\, \Pi_- = \cp^{(m)}_- \cp^{(n)}_+ ,\nn \\
\Pi_0 &=& \fr{\cp^{(m)}_0 -\cp^{(n)}_0 }{2},
 \eea will also
satisfy an polynomial algebra $m+n+1$ if $\Pi   = \fr{\cp^{(m)}_0
+\cp^{(n)}_0 }{2}$ is a constant of motion.

Proof:\\ \bea \lsb \Pi_0 , \Pi_{\pm} \rsb &=& \fr{1}{2} \lsb
\cp_0^{(m)} + \cp_0^{(n)} ,\cp^{(m)}_{\pm} \cp^{(n)}_{\pm} \rsb
\nn \\
                                      &=& \pm \cp^{(m)}_{\pm} \cp^{(n)}_{\pm} \nn \\
                                      &=& \pm \Pi_{\pm}
\eea \bea \lsb \Pi_+ , \Pi_{-} \rsb   &=& \lsb \cp^{(m)}_{+}
\cp^{(n)}_{+}  , \cp^{(m)}_{-} \cp^{(n)}_{-} \rsb  \\ \nn
                                      &=& \cp^{(m)}_{+} \cp^{(m)}_{-} f_{n} (\cp^{(n)}_0 ) + \cp^{(n)}_{-} \cp^{(n)}_{+} f_{m} (\cp^{(m)}_0 )
                                      \nn \\
                                      &=& \lrb C^{(m)} -g_{m+1} (\cp^{(m)}_0 -1) \rrb f_n (\cp^{(n)}_0
                                      ) \\ \nn &+&\lrb C^{(n)} -g_{n+1} (\cp^{(n)}_0 -1) \rrb f_m (\cp^{(m)}_0 )
                                       \\ \nn
\eea  $C^{(m)}$ and $C^{(n)}$ are the Casimir operators. Writing
$\cp_0$ in terms of $\Pi_0 $ and $\Pi$ we get \bea RHS &=&
C^{(m)}\sum _{l=0}^{n} C_l (\Pi_0 -\Pi )^l \\ \nn & -&
\sum_{l=0}^{m+1} \sum_{s=0}^{n}
         a_l c_s (\Pi_0 +\Pi)^l (\Pi_0 -\Pi)^s \nn \\
    &  +& C^{(k)}\sum_{l=0}^{m} d_l (\Pi_0 + \Pi )^l \\ \nn &-& \sum_{l=0}^{n+1} \sum_{s=0}^{m}
        b_l d_s (\Pi_0 -\Pi)^l (\Pi_0 +\Pi)^s \nn \\
    &=& f_{m+n+1} (\Pi_0^{m+n+1} )
\eea $f_{m+n+1} (\Pi_0^{m+n+1} )$ a polynomial in $\Pi_0^{m+n+1}$
of order $m+n+1$ . Using the above construction the tables 1 and 2
give the classification of Quadratic and cubic algebras.\\ {\bf
Classification of Quadratic Algebras}\\
\begin{small}
\begin{tabular}{|l|l|l|l|} \hline
 case& J-S rep. &Algebra&Constants\\
  \hline $Q^-(2)$ & $ Q_0=\frac{1}{2}(J_0-N); $
&$\lsb Q_0,Q_\pm \rsb=\pm Q_{\pm},$ &$ L=\frac{1}{2}(J_0+N)$\\
&$Q_+=J_+a;Q_-=J_-a^{\dag}$&$\lsb
Q_+,Q_-\rsb=-3Q_0^2-(2L-1)Q_0+(J+L(L+1))$&\\\hline $Q^+(2)$ & $
Q_0=\frac{1}{2}(J_0+N); $ &$\lsb Q_0,Q_\pm \rsb=\pm Q_{\pm},$ &$
L=\frac{1}{2}(J_0-N)$\\ &$Q_+=J_+a^{\dag};Q_-=J_-a$&$\lsb
Q_+,Q_-\rsb=-3Q_0^2+(2L+1)Q_0-(J+L(L-1))$&\\\hline $Q^-(1,1)$ & $
Q_0=\frac{1}{2}(K_0-N); $ &$\lsb Q_0,Q_\pm \rsb=\pm Q_{\pm},$ &$
L=\frac{1}{2}K_0+N)$\\ &$Q_+=K_+a;Q_-=K_-a^{\dag}$&$\lsb
Q_+,Q_-\rsb=+3Q_0^2-(2L-1)Q_0+(K+(L(L+1)))$&\\\hline$Q^+(1,1)$ & $
Q_0=\frac{1}{2}(K_0+N); $ &$\lsb Q_0,Q_\pm \rsb=\pm Q_{\pm},$ &$
L=\frac{1}{2}(K_0-N)$\\ &$Q_+=K_+a^{\dag};Q_-=J_-a$&$\lsb
Q_+,Q_-\rsb=-3Q_0^2-(2L-1)Q_0+(K+(L(L+1)))$&\\\hline
\end{tabular}
\end{small}

\newpage

\oddsidemargin -.01cm

{\bf  Classification of Cubic Algebras}\\
\begin{small}
\begin{tabular}{|l|l|l|l|} \hline
 case& J-S rep. &Algebra&Constants\\
  \hline $C^-(11,11)$ & $ C_0=\frac{1}{2}(L_0-M_0) $
&$\lsb C_0,C_\pm \rsb=\pm C_{\pm},$ &$ K=\frac{1}{2}(L_0+M_0)$\\
&$C_+=\mu L_+M_-;C_-=L_-M_+$&$\lsb
C_+,C_-\rsb=-4\mu^2C_0^3-C_0(K^2\mu^2-\sigma)+\lambda K$&
\\\hline $C^+(11,11)$ & $ C_0=\frac{1}{2}(L_0+M_0); $ &$\lsb
C_0,C_\pm \rsb=\pm C_{\pm},$ &$ K=\frac{1}{2}(L_0-M_0)$\\ &$C_+=
L_+M_+;C_-=M_-L_-$& $\lsb
C_+,C_-\rsb=-4\mu^2C_0^3+C_0(K^2\mu^2-\sigma)+\lambda K$&
\\ \hline
$C^-(2,2)$ & $ C_0=\frac{1}{2}(J_0-P_0); $ &$\lsb C_0,C_\pm
\rsb=\pm C_{\pm}$ &$ K=\frac{1}{2}(J_0+P_0)$ \\
&$C_+=J_+P_-;C_-=J_-P_-$&$\lsb
C_+,C_-\rsb=+4C_0^3-(4K^2+\sigma)C_0+\lambda K$&\\\hline$C^+(2,2)$
& $ C_0=\frac{1}{2}(J_0+P_0); $ &$\lsb C_0,C_\pm \rsb=\pm
C_{\pm},$ &$ K=\frac{1}{2}(J_0-P_0)$\\
&$C_+=J_+P_+;C_-=J_-P_-$&$\lsb
C_+,C_-\rsb=4C_0^3-(2K^2-\frac{1}{2}\sigma)C_0+\lambda K$ \\
\hline
 $C^-(2,11)$ & $ C_0=\frac{1}{2}(J_0-L_0); $
&$\lsb C_0,C_\pm \rsb=\pm C_{\pm},$ &$ K=\frac{1}{2}(J_0+L_0)$\\
&$C_+= J_+L_-;C_-=J_-L_+$&$\lsb
C_+,C_-\rsb=4C_0^3-4C_0(K^2)+2(J+C_1)$&\\\hline $C^+(2,11)$ & $
C_0=\frac{1}{2}(L_0+J_0); $ &$\lsb C_0,C_\pm \rsb=\pm C_{\pm},$ &$
K=\frac{1}{2}(J_0-L_0)$\\ &$C_+= L_+J_+;C_-=J_-L_-$&$\lsb
C_+,C_-\rsb=4C_0^3-4C_0 K^2+2(J-C_1)$&\\\hline $C^-(q_-1,h)$ & $
C_0=\frac{1}{2}(Q_0+N); $ &$\lsb C_0,C_\pm \rsb=\pm C_{\pm},$ &$
K=\frac{1}{2}(Q_0-N)$\\ &$C_+=Q_+a^{\dag};C_-=Q_-a$&$\lsb
C_+,C_-\rsb=-4C_0^3-(6K^2+3(L+1))C_0^2+B C_0+D $&
\\ \hline$C^-(q_-1,h)$ & $ C_0=\frac{1}{2}(Q_0-N); $ &$\lsb
C_0,C_\pm \rsb=\pm C_{\pm}$ &$ K=\frac{1}{2}(Q_0+N)$\\
&$C_+=Q_+a;C_-=Q_-a^{\dag}$&$\lsb
C_+,C_-\rsb=4C_0^3-(6K-L-2)C_0^2+ EC_0+F $& \\\hline
 $C^+(q+1,h)$ & $
C_0=\frac{1}{2}(Q_0+N); $ &$\lsb C_0,C_\pm \rsb=\pm C_{\pm},$ &$
K=\frac{1}{2}(Q_0-N)$\\ &$C_+=Q_+a^{\dag};C_-=Q_-a$&$\lsb
C_+,C_-\rsb=-4C_0^3-(6K+3L)C_0^2+B C_0+D $& \\\hline$C^-(q_+1,h)$
& $ C_0=\frac{1}{2}(Q_0-N); $ &$\lsb C_0,C_\pm \rsb=\pm C_{\pm},$
&$ K=\frac{1}{2}(Q_0+N)$\\ &$C_+=Q_+a;C_-=Q_-a^{\dag}$&$\lsb
C_+,C_-\rsb=4C_0^3-(6K-6L-2)C_0^2+ EC_0+F $&\\\hline
\end{tabular}
\end{small}

\oddsidemargin 1cm

\section{ Construction Of Representations}

Since we have constructed a higher order algebra from a  lower
algebra the representation of the higher order algebra can also be
constructed from the lower order algebra. For example, let $ \mid
m,\lambda^{(m)}> $ and  $ \mid n,\lambda^{(n)}> $ be the basis
states for the irreducible unitary representations of the
polynomial algebras $(\cp^{(m)}_0 , \cp_{\pm}^{(m)})$ and
$(\cp^{(n)}_0 , \cp^{(n)}_{\pm})$ respectively, where
$\lambda^{(m)}$ and $\lambda^{(n)}$ $\in \bf{R}$, labeling the
unitary irreducible  representations,\large \bea \cp_0^{(m)} \lmd
q, \lambda^{(m)} \rra &=& q+ \lambda^{(m)} \lmd q, \lambda^{(m)}
\rra \nn \\ \cp_{+}^{(m)} \lmd q, \lambda^{(m)} \rra &=&
\sqrt{t^{\lambda}_{q+1}} \lmd q+1, \lambda^{(m)} \rra \nn \\
\cp_{-}^{(m)} \lmd q, \lambda^{(m)} \rra &=&
\sqrt{t^{\lambda}_{q}} \lmd q-1, \lambda^{(m)} \rra \eea

\bea \cp_0^{(n)} \lmd s, \lambda^{(n)} \rra &=& s+ \lambda^{(n)}
\lmd s, \lambda^{(n)} \rra \nn \\ \cp_{+}^{(n)} \lmd s,
\lambda^{(n)} \rra &=& \sqrt{d^{\lambda}_{s+1}} \lmd s+1,
\lambda^{(n)} \rra \nn \\ \cp_{-}^{(n)} \lmd s, \lambda^{(n)} \rra
&=& \sqrt{d^{\lambda}_{s}} \lmd s-1, \lambda^{(n)} \rra \eea where
$q,s = 0,\pm 1 , ...$

The fact that $\Pi $ is a constant over the product state $\lmd q,
\lambda^{(m)} \rra *\lmd s, \lambda^{(n)} \rra $ gives the
condition,
\be
2\Pi = \lambda^{(m)} + \lambda^{(n)} + q+s \ee If we impose the
above constraint in the product states the following
representations are possible:\large \bea \Pi_0^{(m)} \lmd q,
\lambda^{(m)},\lambda^{(n)},\Pi \rra &=& q+ \lambda^{(m)} -\Pi
\lmd q, \lambda^{(m)},\lambda^{(n)},\Pi \rra  \nn \\
\Pi_{+}^{(m)}\lmd q, \lambda^{(m)},\lambda^{(n)},\Pi \rra  &=&
\sqrt{t^{\lambda^{(m)}}_{q+1}d^{\lambda^{(n)}}_{2\Pi
-\lambda^{(m)} -\lambda^{(n)} +1-q }}\lmd q+1,
\lambda^{(m)},\lambda^{(n)},\Pi \rra  \nn \\ \Pi_{-}^{(m)}\lmd q,
\lambda^{(m)},\lambda^{(n)},\Pi \rra  &=&
\sqrt{t^{\lambda^{(m)}}_{q}d^{\lambda^{(n)}}_{2\Pi -\lambda^{(m)}
-\lambda^{(n)} -q }}\lmd q-1, \lambda^{(m)},\lambda^{(n)},\Pi \rra
\eea The dimension of the representations are decided by the
condition \bea t_{q}^{\lambda^{(m)}} d^{\lambda^{(n)}}_{2\Pi
-\lambda^{(m)} -\lambda^{(n)} -q} &\ge& 0 \nn \\ 2\Pi
-\lambda^{(m)} -\lambda^{(n)} -q \ge 0 \eea \\

 As an example of our method we construct the representation of $Q^{-}(1,1)$, given by \bea Q_0 & = &
 = \fr{1}{2} \lrb K_0 - a_3^\da a_3 \rrb\,, \nn \\
Q_+ & = &  K_+ a_3\,, \qquad Q_- = Q_+^\da =  K_- a_3^\da\,, \nn
\\ \ck & = & \frac{1}{4}
  \lsb 1 - \lrb a_1^\da a_1 - a_2^\da a_2 \rrb^2 \rsb = K^2\,, \nn \\
\cl & = &
      \frac{1}{2} \lrb K_0 + a_3^\da a_3 \rrb\,,
\nn \eea where $\lrb K_0 , K_+ , K_- \rrb$ generate $su(1,1)$ with
$K^2$ as the Casimir operator.  $\lrb Q_0 , Q_+, Q_- \rrb$
generate a quadratic algebra: \bea \lsb \ck , \cl \rsb & = & 0\,,
\quad \lsb \ck , Q_{0,\pm} \rsb = 0\,, \quad \lsb \cl , Q_{0,\pm}
\rsb = 0\,, \nn \\ \lsb Q_0 , Q_\pm \rsb & = & \pm Q_\pm\,,
\nn\\\quad \lsb Q_+ , Q_- \rsb &= &3Q_0^2 + (2\cl-1) Q_0 \nn \eea
with $\ck$ and $\cl$ taking constant values in any irreducible
representation.  The Casimir operator of this algebra is given by
\be \cc = Q_+ Q_- + Q_0^3 + (\cl - 2) Q_0^2
        + \lrb \ck - \cl^2 -2\cl + 1 \rrb Q_0\,,
\ee The condition that $\ck$ and $\cl$ take constant values in an
irreducible representation fixes the basis to be the set of
three-mode Fock states \be \lmd k , l , n \rra = \lmd n, n+2k-1,
2l-k-n \rra\,, \quad
 n = 0, 1, 2,\,\ld\,, (2l-k)\,,
\nn \ee with \be 2l-k = 0,1,2,\,\ld\,, \qquad k = 1/2, 1,
3/2,\,\ld\,, \ee and \be \ck \lmd k , l , n \rra = k(1-k) \lmd k ,
l , n \rra\,, \quad \cl \lmd k , l , n \rra = l \lmd k , l , n
\rra\,. \ee The basis states  carry the $(2l-k+1)$-dimensional
unitary irreducible representation of the quadratic algebra which
can be labeled by the values of the pair $(k,l)$. Explicitly, the
$(k,l)$-th representation is\,: \bea Q_0
\lmd k , l , n \rra & = & (k-l+n) \lmd k , l , n \rra\,, \nn \\
Q_+ \lmd k , l , n \rra & = &
  \sqrt{(n+1)(n+2k)(2l-n-k)}\,\lmd k , l , n+1 \rra\,, \nn \\
Q_- \lmd k , l , n \rra & = &
  \sqrt{n(n+2k-1)(2l-n-k+1)}\,\lmd k , l , n-1 \rra\,. \nn \\
\ck \lmd k , l , n \rra & = & k(1-k) \lmd k , l , n \rra\,, \quad
\cl \lmd k , l , n \rra = l \lmd k , l , n \rra\,, \nn \\
 &  & \qquad \qquad n = 0,1,2,\,\ld\,,(2l-k)\,.
\lb{mrkl} \eea The Casimir operator has the value $\lrb l^3 +
(l+1)[k(1-k)-1] + 1 \rrb$ in this representation.  \\
 For example, for each value of $k =
1/2,1,3/2,\ld\,,$ there is a $2$-dimensional representation of the
algebra  given by \bea Q_0 & = & \fr{1}{2} \lrb \begin{array}{cc}
                     k-1 & 0   \\
                     0   & k+1 \end{array} \rrb\,, \quad
Q_+ = \lrb \begin{array}{cc}
                   0 & 0 \\
           \sqrt{2k} & 0 \end{array}\rrb\,, \quad
Q_- = \lrb \begin{array}{cc}
          0 & \sqrt{2k}  \\
          0 & 0          \end{array} \rrb\,, \nn \\
\ck & = & k(1-k)\,, \quad \cl = l = \frac{1}{2}(k+1)\,,\\ \nn
\quad \cc &= &\frac{1}{8}\lrb -3k^3 - 5k^2 + 11k - 3 \rrb\,, \nn
\eea If we make the association \be \lmd k , l , n \rra \lra
\fr{z_2^{2k-1} z_3^{2l-k} \lrb z_1z_2/z_3 \rrb^n}
                            {\sqrt{n!(n+2k-1)!(2l-k-n)!}}\,.
\ee Since $k$ and $l$ are constants for a given representation we
can take
\be
\phi_{k,l,n}(z) = \fr{z^n}{\sqrt{n!(n+2k-1)!(2l-k-n)!}}\,, \quad
  n = 0,1,2,\,\ld\,,(2l-k)\,,
\ee With this we get the Fock-Bargmann representation of $Q(1,1)$

\be
Q_0 = z \ddz + k - l\,, \quad Q_+ = -z^2 \ddz + (2l-k)z\,, \quad
Q_- = z \ddzt + 2k \ddz\,, \ee

Using this method one can get the Bargmann schwinger
representations of all the 16 algebras given in tables 1 and 2.
These are useful in construction of Barut girardello coherent
states of polynomial algebras as discussed in ref \cite{vsk}.

\section{ Polynomial Algebras  In Supersymmetric Quantum Mechanics}Recent work on
supersymmetric methods  for the construction of conditionally
exactly solvable problems has shown that these systems have a
non-linear algebraic structure, with the symmetry algebra being a
polynomial algebra \cite{junker , roy,  hussin}. Symmetry algebra
of SUSY partner of Linear Oscillator has a Quadratic algebra
structure, while the SUSY partner of the radial oscillator has a
Cubic algebra structure . These algebras belong to our
classification. to illustrate this consider the standard
construction  of the Susy partner  of the Hamiltonian. Given the
hamiltonian
\be
H_0=-\frac{1}{2}(\frac{d^2}{dx^2}+ V_{0}(x) \ee we construct its
supersymmetric partner
\be
H_{-}=-\frac{1}{2}(\frac{d^2}{dx^2}+ V_{1}(x)\ee by defining
\be
A=\frac{1}{\sqrt{2}}(-\frac{d}{dx}+W(x))\ee such that
\be
H_1A^{\dag}=A^{\dag}H_0 \ee

One makes the ansatz $W(x)=\phi(x)+f(x)$ , where$f(x)$ satisfies
Riccati eqn. and $\phi(x)$ is the shape invariant susy potential
and  \be f^2(x)+2\phi(x)f(x)+f'(x)=2(\epsilon-1)\ee. For the case
of the  harmonic oscillator  $\phi(x)=x$, Junker and Roy construct
Ladder operators for $H_-$, $B=A^{\dag}aA\;\;\;
B^{\dag}=A^{\dag}a^{\dag}A$ which along with $H_-$ satisfy the
algebra: \be\lsb H_-,B\rsb=-B\,\,\,\lsb
H_-,B^{\dag}\rsb=-B^{\dag}\,\,\,\lsb
B^{\dag},B\rsb=-3H_{-}^{2}+4\epsilon H_{-}-\epsilon^2\ee. This
corresponds to algebra $Q^{-}(1,1)$ and the corresponding
constituent Lie algebras  are the Heisenberg algebra generated by
$A,A^\dag,A^{\dag}A$ and the SU(1,1) algebra generated by
$K_+=A^{\dag}a^{\dag},K_-=Aa,K_0=\frac{1}{2}(a^{\dag}a+A^{\dag}A+1)$.

Similarly for the radial oscillator one has
$\phi(x)=x-\frac{\gamma+1}{x}$. If $c,c^{\dag}$ are the creation
and annihilation  operators for $H_+$  such that
\be
\lsb H_+,c\rsb=-2c\,\,\,\lsb H_+,c^{\dag}\rsb=-2c^{\dag}\,\,\,\lsb
c^{\dag},c\rsb=-4H_{+}+\gamma+\epsilon-\frac{3}{2}\ee then for
$H_-$ the operators $D=A^{\dag}cA\;, D^{\dag}=A^{\dag}c^{\dag}A$
together with $H_-$ satisfy the algebra: \be\lsb
H_-,D\rsb=-D\,\,\,\lsb H_-,D^{\dag}\rsb=-D^{\dag}\,\,\,\lsb
D^{\dag},D\rsb=-8H_{-}^{3}-12(\gamma+\epsilon
+\frac{1}{2})H_{-}^2+4(\epsilon^2+\epsilon+1+2\epsilon \gamma)\ee
This corresponds to algebra $C^{-}(q_-1,h)$ of our classification

This method has been generalized by Hussin, Fernandez and Nieto
using the method of intertwining operators to construct higher
order SUSY potentials all of which belong to our classification
and therefore the coherent states and representations can be read
off the list.

\section{Further Applications}
We now consider some interesting applications of polynomial
algebras to systems which are linked to supersymmetric quantum
mechanics.
\subsection{ Quadratic oscillator}
An interesting possibility is suggested by the structure of the
algebra $Q^-(1,1)$.  Let us define
\be
N = Q_0\,, \quad A = \frac{1}{\sqrt{\cl(\cl+1)-\ck}}\,Q_-\,, \quad
A^\da = \frac{1}{\sqrt{\cl(\cl+1)-\ck}}\,Q_+\,. \ee Then the
algebra $Q^-(1,1)$ becomes \bea \lsb N , A \rsb & = & -A\,, \quad
\lsb N , A^\da \rsb = A^\da\,, \nn \\ \lsb A , A^\da \rsb & = & 1
- \fr{2\cl-1}{\cl(\cl+1)-\ck}\,N
                            - \fr{3}{\cl(\cl+1)-\ck}\,N^2\,.
\lb{qo} \eea We may consider this as the defining algebra of a
quadratic oscillator, corresponding to a special case of the
general class of deformed oscillators :
\be
\lsb N , A \rsb = -A\,, \quad \lsb N , A^\da \rsb = A^\da\,, \quad
\lsb A , A^\da \rsb = F(N)\,. \ee The quadratic oscillator
(\ref{qo}) belongs to the class of generalized deformed
parafermions \cite{Q}.  It should be interesting to study the
physics of assemblies of quadratic oscillators.  In fact, the
canonical fermion, with
\be
N = \lrb \ba{cc}
         0 & 0  \\
         0 & 1 \ea \rrb\,, \quad
f = \lrb \ba{cc}
         0 & 1 \\
         0 & 0 \ea \rrb\,, \quad
f^\da = \lrb \ba{cc}
          0 & 0  \\
          1 & 0  \ea \rrb\,,
\ee is a quadratic oscillator!  Observe that
\be
\lsb N , f \rsb = -f\,, \quad \lsb N , f^\da \rsb = f^\da\,, \quad
\lsb f , f^\da \rsb = 1 - \half N - \fr{3}{2} N^2\,. \ee

\subsection{ Two Body Calogero-Sutherland Model}
The Hamiltonian of the two body Calogero-Sutherland model is given
by \be H_c = \fr{1}{2} \sum_{i=1}^{2} (\fr{\partial^2}{\partial
x^2 } + \omega ^2 x^{2}_i +\sum_{j<i} \fr{\lambda (\lambda
-1)}{(x_i -x_j )^2 } \ee defining the creation and annihilation
operators, \bea a^{\dagger}_i &=& \fr{1}{\sqrt{2}} (-D_i + \omega
x_i )  \nn
\\ a_i &=& \fr{1}{\sqrt{2}} (D_i + \omega x_i ) \eea
Where $ D_i$ is the Dunkel derivative \be D_i =
\fr{\partial}{\partial x } + \lambda \sum_{i\neq j}\fr{1} {(x_i
-x_j )} (1-\sigma_{ij}) \ee
 In the centre of mass
coordinates defined by \bea y_i    &=& M_{ij} x_j ,\\ M_{ij} &=&
\fr{1}{\sqrt{2}} \left[ \begin{array}{cc} -1 & 1 \\
 1 & 1
\end{array}
\right] \eea the corresponding creation and annihilation operators
are given by \be \tilde{A}_i = M_{ij} a_j , \,\,
\tilde{A}^{\dagger}_i = M_{ij} a^{\dagger}_j \ee Defining:
  \bea A_1 &=& \fr{\tilde{A}_1 }{\sqrt{1+
2\lambda \sigma }} \nn \\ A^{\dagger}_1 &=&
\fr{\tilde{A}^{\dagger}_1 }{\sqrt{1+ 2\lambda \sigma }} \nn \\ A_2
&=& \tilde{A}_2 , \,\, A^{\dagger}_2 =\tilde{ A}^{\dagger} \eea
The operators
 \be C_0 =\fr{1}{2} (A^{\dagger}_1 A_1 -A^{\dagger}_2 A_2 ) C_+
=\fr{1}{2} (A_{1}^{\dagger} A_2 )^2 \ee \be C_- = \fr{1}{2}
(A_{2}^{\dagger} A_1 )^2  J   =  \fr{1}{2} (A^{\dagger}_1 A_1 +
A^{\dagger}_2 A_2 ) \ee  satisfy the cubic algebra \be
 \lsb
C_0 , C_{\pm} \rsb = C_{\pm} \ee \be \lsb C_+ , C_- \rsb = -2
C^{3}_0 + (2C(J)-1 )C_0 \ee.

The Hamiltonian in terms of the operator can written as
 \be
H= 2\lambda \sigma ) C_0 + (1+\lambda \sigma )(2J +1) \ee

 This is significantly different from
the algebra generated by Vinet et. al. \ref{vinet}, as the
Hamiltonian is one of the generators. Here C(J)=J(J+1) is the
Casimir of the underlying $SU(2)$ algebra. Note that the above
algebra has a one mode $SU(2)$ realization, given by $
 C_0 = J_0
, C_+ = \fr{1}{2} J^2_+ , C_- = \fr{1}{2} J^2_-
 $

\section{Conclusion} We have seen that polynomial algebras emerge
as the dynamical symmetry , invariance and spectrum generating
algebras of many interesting physical systems . Among these, a
special position is occupied by three dimensional polynomial
algebras with a coset structure. Thus, a systematic study and the
proper classification of these algebras and their irreducible
representations was warranted. Such a comprehensive study has been
carried out in detail.
\section{Acknowledgements:}
I would like to thank Profs. L.M.Nieto and J. Negro for their
hospitality in Valladolid ,Spain. This work was done in
collaboration with Prof. R. Jagannathan, IMSC, Chennai and Dr.V.
Sunilkumar, IOP, Bhubaneshwar. I would also like to put on record
my thanks to Prof. U. Sukhathme and Prof. A. Khare whose talks  at
Hyderabad  were the inspiration of this work. many thanks to Prof.
C. Mukku U.G.C. Scientist `C' (professors grade) for his untiring
help with the manuscript. Financial thanks to DST, INSA and the
University of Hyderabad for travel support.

\end{document}